\documentclass[10pt,letterpaper]{article}

\usepackage[T1]{fontenc}
\usepackage[utf8]{inputenc}
\usepackage{newtxtext,newtxmath}
\usepackage{microtype}
\usepackage[letterpaper,top=0.68in,bottom=0.72in,left=0.74in,right=0.74in,
            headheight=15pt,headsep=16pt,footskip=26pt]{geometry}
\usepackage{graphicx}
\usepackage{xcolor}
\usepackage{tikz}
\usetikzlibrary{arrows.meta,positioning}
\usepackage{booktabs}
\usepackage{float}
\usepackage{array}
\usepackage{enumitem}
\usepackage{amsmath}
\usepackage{titlesec}
\usepackage{fancyhdr}
\usepackage[font=small,labelfont={bf,rm},labelsep=period]{caption}
\usepackage[numbers,square,sort&compress]{natbib}
\usepackage[most]{tcolorbox}
\usepackage{hyperref}

\definecolor{Cobalt}{HTML}{063166}
\definecolor{Teal}{HTML}{5384B2}
\definecolor{Ink}{HTML}{172539}
\definecolor{Subtle}{HTML}{58697C}
\definecolor{SoftRule}{HTML}{D9E3ED}
\definecolor{Pale}{HTML}{F1F6FB}
\definecolor{PaleBlue}{HTML}{E7F0F8}
\hypersetup{colorlinks=true,linkcolor=Cobalt,citecolor=Cobalt,urlcolor=Cobalt,
  pdftitle={APCL Research White Paper Template},
  pdfauthor={Advanced Propulsion Concepts Lab (APCL)}}

\newcommand{\papertitle}{Physics-based Tabulation Approach \\for Rotating Detonation Engines}

\newcommand{\paperauthors}{%
  Noah Schwalb\textsuperscript{1},\quad
  Vansh Sharma\textsuperscript{1},\quad
  Shivank Sharma\textsuperscript{1},\quad
  Venkat Raman\textsuperscript{1}}
\newcommand{\paperdate}{November 2026}
\newcommand{\paperaffiliations}{%
  \textsuperscript{1}Department of Aerospace Engineering, University of Michigan, Ann Arbor, 48109, USA\\[2pt]}

\newcommand{\papercorrespondence}{Noah Schwalb (\href{mailto:author@example.org}{nschwalb@umich.edu})}

\color{Ink}
\setlist[itemize]{topsep=3pt,itemsep=3pt,parsep=0pt,leftmargin=1.35em,
  label={\textcolor{Cobalt}{\large\textbullet}}}
\titleformat{\section}{\rmfamily\bfseries\large\color{Ink}}
  {\textcolor{Cobalt}{\thesection}}{0.70em}{}
\titleformat{\subsection}{\rmfamily\bfseries\normalsize\color{Ink}}
  {\textcolor{Cobalt}{\thesubsection}}{0.70em}{}
\titlespacing*{\section}{0pt}{17pt plus 2pt minus 1pt}{6pt}
\titlespacing*{\subsection}{0pt}{11pt plus 1pt minus 1pt}{4pt}
\fancypagestyle{apcl}{%
  \fancyhf{}%
  \fancyhead[L]{\rmfamily\fontsize{8}{10}\selectfont\bfseries\color{Ink}APCL}%
  \fancyhead[R]{\rmfamily\fontsize{7.5}{9}\selectfont\color{Subtle}Tabulation for RDEs}%
  \fancyfoot[R]{\rmfamily\fontsize{8.5}{10}\selectfont\bfseries\color{Cobalt}\thepage}%
}
\fancypagestyle{apclfirst}{%
  \fancyhf{}%
  \fancyfoot[L]{\rmfamily\fontsize{7.6}{9.2}\selectfont\color{Subtle}\textcopyright\ 2026 All rights reserved by the Author and Advanced Propulsion Concepts Lab.}%
  \fancyfoot[R]{\rmfamily\fontsize{8.5}{10}\selectfont\bfseries\color{Cobalt}\thepage}%
}
\newtcolorbox{abstractpanel}{
  enhanced,breakable,boxrule=0pt,colback=Pale,
  borderline west={2.2pt}{0pt}{Cobalt},
  sharp corners,left=13pt,right=12pt,top=11pt,bottom=10pt,
  before skip=0pt,after skip=8pt
}
\newcommand{\minilabel}[1]{{\rmfamily\fontsize{8.5}{10}\selectfont\bfseries\color{Cobalt}\MakeUppercase{#1}}}

\begin{document}
\thispagestyle{apclfirst}

\noindent\begin{minipage}[c]{0.60\linewidth}
  \includegraphics[width=1.69in]{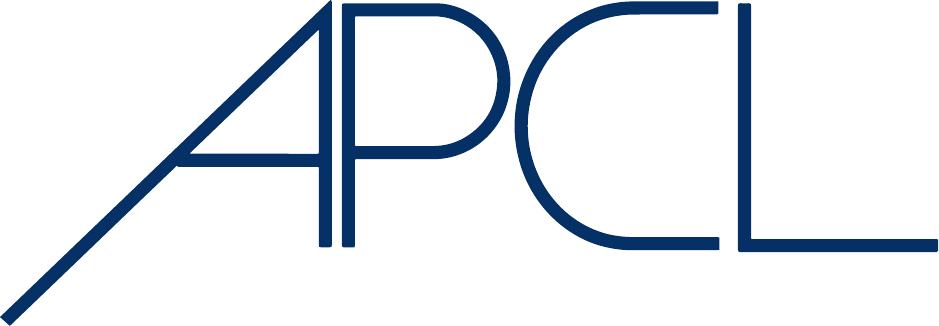}\par\vspace{2pt}
\end{minipage}\hfill\begin{minipage}[c]{0.37\linewidth}
  \raggedleft\rmfamily\fontsize{7.6}{10.2}\selectfont
  \color{Subtle}RESEARCH SERIES\quad$\bullet$\quad \paperdate
\end{minipage}

\vspace{2pt}
\noindent\textcolor{Cobalt}{\rule{1.69in}{1.6pt}}%
\textcolor{SoftRule}{\rule{\dimexpr\linewidth-0.62in\relax}{0.8pt}}
\vspace{10pt}

{\rmfamily\bfseries\fontsize{23}{27}\selectfont\color{Ink}\papertitle\par}
\vspace{5pt}
\vspace{12pt}
{\rmfamily\fontsize{9.9}{12.4}\selectfont\bfseries\color{Ink}\paperauthors\par}
\vspace{4pt}
{\rmfamily\fontsize{8.9}{12.2}\selectfont\color{Subtle}\paperaffiliations\par}
\vspace{12pt}

\begin{abstractpanel}
  \minilabel{Abstract}\par\vspace{3pt}
High-fidelity simulations of rotating detonation engines remain computationally expensive due to the strong coupling between compressible flow dynamics and stiff chemical kinetics. The coexistence of deflagrative and detonative combustion, along with shock interactions and rapid thermodynamic changes, further challenges conventional reduced-order chemistry models. This work presents a multi-mode tabulation framework that replaces detailed chemical kinetic integration with precomputed thermochemical states and source terms while accounting for variations in the local thermodynamic state. The key novelty lies in a unified treatment of distinct combustion regimes, wherein homogeneous reactor solutions describe deflagrative combustion and post-detonation reactions up to the Chapman-Jouguet (CJ) point, while an equilibrium isentropic expansion model captures supersonic relaxation beyond the CJ state. The framework is evaluated against detailed chemistry across progressively complex configurations, including homogeneous ignition, one-dimensional detonations, and a two-dimensional rotating detonation engine (RDE). The tabulated model accurately reproduces ignition delays and captures the initiation and sustained propagation of detonation waves across varying mixture compositions. Furthermore, the framework reconstructs the complete thermodynamic evolution along an RDE streamline, including post-injection expansion, detonation, post-detonation expansion, and oblique shock interactions, directly from conserved flow variables. These results demonstrate the capability of multi-mode tabulation to represent coupled combustion, shock, and expansion processes within a single thermodynamically responsive framework, offering a promising route toward computationally efficient simulations of detonative combustion across a broad range of applications.

  \newcommand{\paperkeywords}{Rotating Detonation Engine; Detonation; Tabulation; Reduced-Order Model; Non-premixed Combustion}
  
  \par\vspace{8pt}
  {\color{SoftRule}\rule{\linewidth}{0.55pt}}\par\vspace{2pt}
  {\rmfamily\fontsize{8.5}{11.5}\selectfont\color{Subtle}%
    \textbf{\textcolor{Ink}{Keywords}}\quad\paperkeywords\par
    \vspace{3pt}
    \textbf{\textcolor{Ink}{Corresponding author}}\quad\papercorrespondence\par
  }
\end{abstractpanel}

\section{Introduction} \label{sec:intro}

Detonation-based combustion systems offer a potential route to pressure-gain propulsion and energy conversion, but their predictive simulation remains challenging because chemical reactions interact with strongly compressible, unsteady flow~\cite{wolanski_detonative_propulsion,kailasanath_propulsion_review,raman_nonidealities_rde}. In a rotating detonation engine (RDE), one or more detonation waves propagate through a continuously replenished reactant layer while interacting with fuel injection, mixing, shocks, and the expansion of combustion products. In practical RDE, incomplete mixing can also support deflagrative (parasitic) burning ahead of the detonation, altering the thermochemical state of the reactants subsequently consumed by the wave and, consequently, wave dynamics and combustor performance \cite{raman_nonidealities_rde,Sato2023Boundary,Keskinoz2026Partitioned}. A useful chemistry model must therefore describe both deflagrative and detonative heat release.

Predictive RDE simulations are computationally demanding because short detonation induction and reaction timescales coexist with much longer mixing, refill, and flow-development timescales \cite{raman_nonidealities_rde,Barwey2025Timescales}. In addition to mesh and time-integration costs, finite-rate chemistry requires repeated evaluation of nonlinear source terms and integration of stiff kinetic systems. Detailed-chemistry simulations of a full-scale RDE have required approximately $10^{6}$ CPU core-hours per configuration \cite{Sato2023Boundary}. Reducing the cost of source-term evaluation is therefore important for practical parametric studies, although it does not eliminate the cost of resolving the flow.

Reduced chemical mechanisms decrease kinetic complexity but are generally optimized for a restricted range of conditions or target quantities \cite{LuLaw2009Chemistry}. More aggressive approximations replace the resolved detonation reaction zone with prescribed wave and heat-release models. For example, Ref. \cite{Keskinoz2026Partitioned} separated upstream deflagrative heat release from detonation heat release and reported a 19.7-fold reduction in computational cost relative to a detailed-chemistry RDE calculation; the detonation speed, however, was specified from experiment. Full-scale gas-phase RDE simulations with detailed chemistry can require thousands of CPU cores and weeks of computation, even with highly optimized, scalable finite-volume solvers~\cite{numericsHMM, carreon2025gpu}, to reach statistical stationarity and accumulate converged statistics~\cite{powers2026impact_aiaa, powers2026impact}. Chemistry tabulation offers a complementary route by replacing repeated kinetic integration with retrieval of precomputed thermochemical states and source terms \cite{Pope1997ISAT,VanOijen2016FGM, sharma2024stabilization, zhang2025high}.

Tabulated flamelet chemistry has been widely used in low-mach applications~\cite{zhang2025three, sharma2024stabilization, zhang2025high}, and further extended to scramjet simulations \cite{Oevermann2000Scramjet,Torrez2011Scramjets,Pecnik2012HyShot}. Compressible flamelet/progress-variable (FPV) formulations are primarily based on the flamelet framework \cite{IhmePitsch2008Extinction} and account for variations in transported energy, density, and temperature when reconstructing thermochemical states and reaction rates \cite{Vicquelin2011Coupling,Saghafian2015Compressible,Saghafian2015HIFiRE}. However, successful application to supersonic combustion does not establish accuracy for detonations: shock compression and subsequent expansion can move the local state away from the reference flamelet manifold, compromising source-term predictions \cite{Baumgart2025Assessment}. Detonation-oriented extensions include Zel'dovich--von Neumann--D\"oring (ZND)-informed corrections to premixed-flamelet source terms for RDEs \cite{Shunn2023ExtendedFPV,Senga2024ExtendedFPV} and a manifold constructed from ZND solutions \cite{Baumgart2025Detonation}. The latter reduces the cost of detonation chemistry but does not, by construction, ensure fidelity for slower deflagrative burning.

Recent studies have emphasized thermodynamically consistent coupling between compressible flow solvers and chemistry manifolds \cite{CisnerosGaribay2024Coupling,Boerchers2026Consistent}. For a model spanning deflagrations and detonations, source terms must remain responsive to local pressure, temperature, and composition across compression, expansion, and flame propagation. Increasing manifold dimensionality can improve state-space coverage but also raises storage and interpolation costs \cite{Lacey2021Adaptive}. These competing requirements motivate a compact, thermodynamically responsive tabulation strategy for reducing chemical source-term evaluation costs in high-speed reacting flows. 

The present study investigates a \emph{tabulation} approach as a computationally efficient approximation to finite-rate chemical source-term evaluation for both deflagrations and detonations. The objective is to reduce the expense of repeated reactive source-term calculations while retaining sufficient sensitivity to the evolving thermochemical state to describe the distinct reaction pathways and timescales associated with the two combustion regimes. The proposed framework is intended to be evaluated against detailed-chemistry reference solutions in terms of source-term fidelity, heat-release behavior, and computational cost, with particular attention to its suitability for high-speed applications such as RDEs. The tabulation formulation and its construction procedure are described in the following section.

\section{Methodology}
\label{sec:Method}

\subsection{Governing Equations}\label{subsec:NSeq}

The model is presented using the governing equations for a multi-component reacting fluid~\cite{numericsHMM} transformed into $Z-C$ space (see Appendix for derivation)

\begin{equation}
    \frac{\partial \rho}{\partial t} + \frac{\partial (\rho u)}{\partial x_j} = 0
\end{equation}

\begin{equation}
    \frac{\partial (\rho u_j)}{\partial t} + \frac{\partial (\rho u_i u_j)}{\partial x_j}+\frac{\partial p}{\partial x_j} - \frac{\partial \tau_{ij}}{\partial x_j} = 0
\end{equation}

\begin{equation}
    \frac{\partial (\rho E)}{\partial t}
    + \frac{\partial (\rho u_j E)}{\partial x_j}
    + \frac{\partial (u_j p)}{\partial x_j}
    - \frac{\partial (u_i \tau_{ij})}{\partial x_j}
    - \frac{\partial}{\partial x_j}
    \left[
        (\alpha + \rho D c_p) \frac{\partial T}{\partial x_j}
        - \rho D \frac{\partial h}{\partial x_j}
    \right] = 0
\end{equation}

\begin{equation}
    \frac{\partial (\rho Z)}{\partial t} + \frac{\partial (\rho u_j Z)}{\partial x_j} - \frac{\partial}{\partial x_j} \left( \rho D \frac{\partial Z}{\partial x_j} \right) = 0
\end{equation}

\begin{equation}
    \frac{\partial (\rho C)}{\partial t} + \frac{\partial (\rho u_j C)}{\partial x_j} - \frac{\partial}{\partial x_j} \left( \rho D \frac{\partial C}{\partial x_j} \right) = \Omega_C
\end{equation}

\noindent where $\rho$ is the fluid density, $u_j$ is the velocity in direction $j$, $p$ is the thermodynamic pressure, $T$ is the temperature, $\tau_{ij}$ is the viscous stress tensor, $E$ is the total energy, $c_p$ is the mass-specific specific heat at constant pressure of the mixture, $Z$ is the elemental Bilger mixture fraction \cite{bilger1976structure}, and $C$ is the reaction progress variable. The relevant transport coefficients are thermal conductivity $\alpha$, mass diffusivity $D$, and viscosity $\mu$ which is used to evaluate the viscous stress tensor.

\begin{align}
    \tau_{ij} = -\frac{2}{3} \mu \frac{\partial u_m}{\partial x_m} \delta_{ij} + \mu \left( \frac{\partial u_j}{\partial x_i} + \frac{\partial u_i}{\partial x_j} \right)
\end{align}

This model contains two components: deflagrative combustion and supersonic expansion. Combustion dominates behind the detonation wave until the Chapman-Jouguet (CJ) point, after which the flow is dominated by supersonic expansion. Therefore, modeling the combustion that occurs in a detonation is not enough to capture the physics of an RDE and the expansion must be explicitly accounted for. This combined model now referred to as the Multi-mode Tabulation for Detonations (MTD) model.

\subsection{Deflagrative Combustion Model}

Integrating a detailed mechanism in every cell dominates the cost of a reacting-flow simulation. That integration is replaced here by a precomputed table of homogeneous-reactor (HR) solutions, following the tabulation procedure of Tang et~al.~\cite{tang2019comprehensive}. The governing assumption is that the chemical time scale is faster than the transport time scale such that the local state is reaction-dominated and can be represented by a spatially uniform reactor at the same thermodynamic condition.

The table is parametrized by three scalars: the mixture fraction $Z$, in Bilger's elemental form~\cite{bilger1976structure}, measures the local proportion of fuel-stream mass and is unchanged by reaction. The mixture enthalpy $h=\sum_k Y_k h_k(T)$, which accounts for both the chemical and the sensible contribution, is unchanged along an adiabatic constant-pressure path. The progress variable $C=Y_{\mathrm{H_2O}}$, in the current case, changes monotonically through the principal heat-release stage. Consequently, a single reactor run has fixed $(Z,h,p)$ and sweeps only $C$, therefore sweeping the initial equivalence ratio populates the $Z$ axis and sweeping the initial temperature populates the $h$ axis~\cite{tang2019comprehensive}. 

Tabulating the progress-variable rate $\dot\omega_C$ directly is inaccurate near ignition, where $\dot\omega_C$ varies by several orders of magnitude within one interval of the $C$ axis, so that a single sampled value substantially overestimates the integrated source~\cite{tang2019comprehensive}. The elapsed reactor time is stored instead. With $t_{\mathrm{HR,Tab}}(C_i)$ being the time at the $i$th node of the $C$ axis, the interval $\Delta t_{\mathrm{HR,Tab}}(C_i)=t_{\mathrm{HR,Tab}}(C_{i+1})- t_{\mathrm{HR,Tab}}(C_i)$ is the time it takes for the real chemistry to cross that bin. During lookup, the solver provides current progress $C(t_0)$ and its own step $\Delta t_{\mathrm{sim}}$. Starting from $C(t_0)$, these bin times are added one bin at a time until the sum reaches $\Delta t_{\mathrm{sim}}$; the progress reached at that point is $C_{\mathrm{target}}$,
The source term follows from its definition:
\begin{flalign}
\footnotesize
\Delta t_{\mathrm{sim}}=\!\!\sum_{C_i=C(t_0)}^{C_{\mathrm{target}}}\!\!
\Delta t_{\mathrm{HR,Tab}}(C_i), \ \
S_{C,r}=\frac{C_{\mathrm{target}}-C(t_0)}{\Delta t_{\mathrm{sim}}} .
\label{eq:src}
\end{flalign}
Thus, the solver thereby advances along the reactor trajectory by the same interval of time as the fluid time step. 

Due to compressibility assumptions within the Navier-Stokes solver used here~\cite{numericsHMM}, pressure becomes a critical thermodynamic variable in the tabulation. Pressure is carried either as a fourth axis, at proportionate cost in memory, or as a single reference pressure $p_0$ with a runtime correction. In the latter formulation, the tabulated $\Delta t_{\mathrm{HR,Tab}}(C_i)$ times are rescaled using the compressible-flamelet relation similar to Cirneros-Garibay et~al.~\cite{cisneros2024consistent} and Saghafian et~al.~\cite{saghafian2015efficient} reduced to its pressure-only part,
\begin{equation}
\Delta t(p,C_i)=\Delta t(p_0,C_i)\left(\frac{p_0}{p}\right)^{a_p(C_i)},
\label{eq:eq26}
\end{equation}
where the exponent $a_p$ is obtained per cell using the least squares method applied to $\ln(\dot\omega/\dot\omega_0)$ against $\ln(\rho/\rho_0)$ and stored with its fit diagnostics. The density and thermal diffusivity are not scaled, but recomputed, $\rho=p \bar{W} /(R_u T)$ and $\alpha=\lambda/(\rho c_p)$. The remaining fields carry no explicit pressure dependence: the thermodynamic data are functions of temperature alone, and at first order in Chapman--Enskog theory $\mu$ and $\lambda$ of a dilute gas are independent of density~\cite{hirschfelder1964molecular}. 

\subsection{Supersonic Expansion Model}
\label{sec:expansion_model}

This model is designed to capture the supersonic expansion that occurs after the CJ point behind a detonation. For any given mixture there is a unique state which is can be identified by $(\rho_{CJ},p_{CJ},Y_{CJ})$ where the subscript $CJ$ corresponds to a state at the CJ point behind a detonation. This point refers to the post-detonation equilibrium condition where heat release $q$ from the reaction zone has choked the flow. Commensal expansion starts from this point, therefore the thermodynamic trajectory behind the CJ point of a detonation lies along the expansion portion of an equilibrium isentrope. This isentrope is therefore parameterized by $(Z,\rho_{CJ},\rho)$ with the same information stored as the deflagrative combustion model.

\subsection{Thermochemical Closure}
\label{sec:thermochemical_closure}

An equation of state is still required to thermochemically close the system, however, mass fraction information is still missing therefore the temperature cannot be inverted from the internal energy and an ideal gas equation of state cannot be used. Only information from the conserved state $\mathbf{U}=[\rho \quad \rho u_j \quad \rho E \quad \rho Z \quad \rho C]$ is available for the thermochemical closure alongside the tabulation, therefore the closure is now of the form

\begin{align}
    (T,p) = \mathcal{F}(\mathbf{U})
\end{align}

\noindent where $\mathcal{F}$ is the closure function that solves for the thermochemical state embedded within the tabulation. In this work, the closure function is a linear search using the parameterization of the deflagrative combustion and supersonic expansion tables to determine the thermodynamic state and transport coefficients from the conserved state.

\section{Results}
\label{sec:results}

\subsection{1D Deflagration Results}\label{sec:deflagrate}

\begin{figure*}[t]
    \centering
    \includegraphics[width=\linewidth]{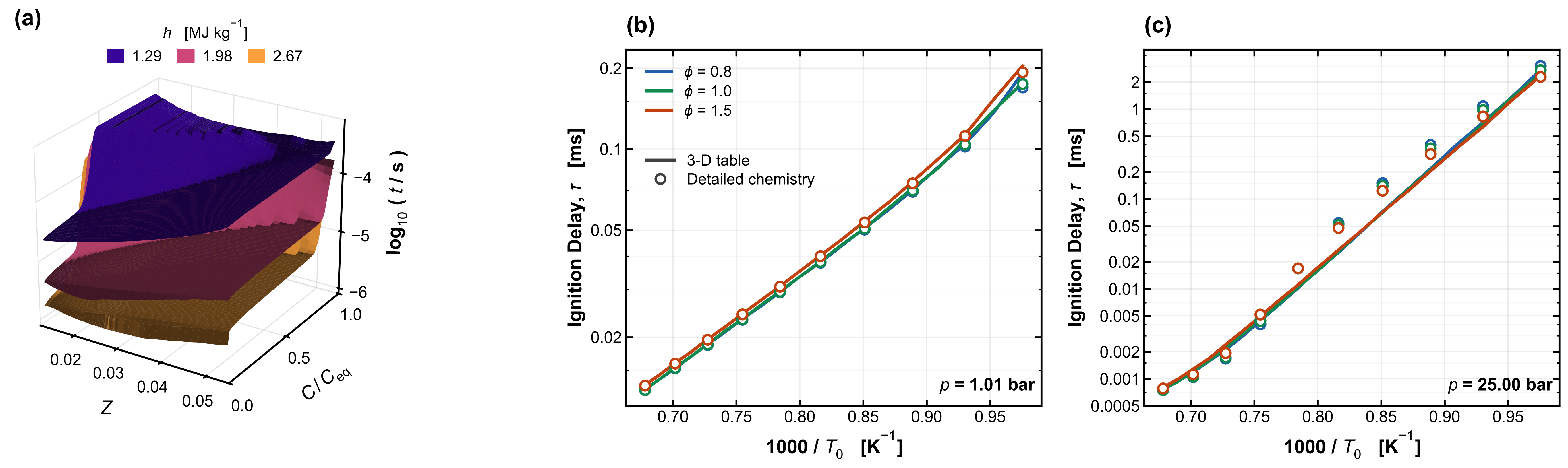}
    \vspace{-5mm}
    \caption{The H\textsubscript{2}--air homogeneous-reactor table results. \textbf{(a)}~The stored manifold at three enthalpies. Each sheet is the elapsed reactor time $t$ over the mixture fraction and the normalized progress variable $C/C_{\text{eq}}$, for $0.013 \le Z \le 0.054$ ($\phi$ from $0.43$ to $1.95$). \textbf{(b,\,c)}~Ignition delay $\tau$, defined as the time at which $\mathrm{d}Y_{\mathrm{OH}}/\mathrm{d}t$ is a maximum, against inverse initial temperature for three equivalence ratios, at the reference pressure $p_0 = 1.013$~bar (b) and at $25$~bar (c). Lines are the table, open symbols direct integration of the mechanism; the legend in (b) applies to both.}
    \label{fig:1d_deflagration}
\end{figure*}

Figure~\ref{fig:1d_deflagration}(a) illustrates the tabulated reaction-time manifold parameterized by mixture fraction ($Z$), normalized progress variable ($C$), and specific enthalpy ($h$). Increasing enthalpy substantially reduces the characteristic reaction time, reflecting the temperature sensitivity of hydrogen oxidation. Normalization by the equilibrium progress variable provides a common representation of the reaction progress across mixture compositions. The stored reaction-time trajectories are subsequently used to reconstruct chemical source terms without direct integration of the detailed kinetic mechanism.

At the reference pressure of 1 atm, the tabulated ignition delays closely reproduce detailed-chemistry predictions over the investigated temperature and equivalence-ratio ranges [Fig.~\ref{fig:1d_deflagration}(b)]. The median relative deviation is below 1\%, with all predictions within 20\% of the reference values. At 25 bar [Fig.~\ref{fig:1d_deflagration}(c)], agreement remains satisfactory above approximately 1200 K, whereas ignition delays are systematically underestimated at lower temperatures. The slight deviation originates primarily from the approximate pressure dependence of the chemical source terms. The density-based power-law correction captures changes in hydrogen chain-branching and termination kinetics near the second explosion limit. These results demonstrate that the three-dimensional manifold accurately represents ignition chemistry at its reference pressure, while its extension to other pressures is limited by the local pressure-scaling approximation. Explicit pressure parameterization may therefore be necessary when the operating conditions span distinct kinetic regimes.

\vspace{-4mm}

\subsection{1D Detonation Initiation}
\label{sec:1d_det}

The tabulation performance is also evaluated in a 1D detonation starting from initiation. The geometry is a tube with length $L = 4.096 \times 10^{-1} \hspace{1mm} m$ with no-slip wall boundary conditions at each end and a stoichiometric $H_2$-air homogeneous mixture with $T = 700 \hspace{1mm} K$ and $p = 1 \hspace{1mm} atm$ throughout the tube. To initiate the detonation, a small kernel of size $L_{ign} = 5 \times 10^{-3} \hspace{1mm} m$ with $T_{ign} = 3,000 \hspace{1mm} K$, $p_{ign} = 12 \hspace{1mm} atm$, and $U_x = 850 \hspace{1mm} m/s$ is placed on the left side of the domain. The ZND induction length of the stoichiometric mixture is $L_{ind} = 2.098 \times 10^{-4} \hspace{1mm} m$, therefore the grid size of $\Delta x = 2.0 \times 10^{-4} \hspace{1mm} m$ allows for one cell per induction length. This is enough to roughly capture the detonation \cite{barwey_detonation_convergence} and is deemed sufficient for this test.

\begin{figure}[h]
    \centering
    \includegraphics[width=0.45\linewidth]{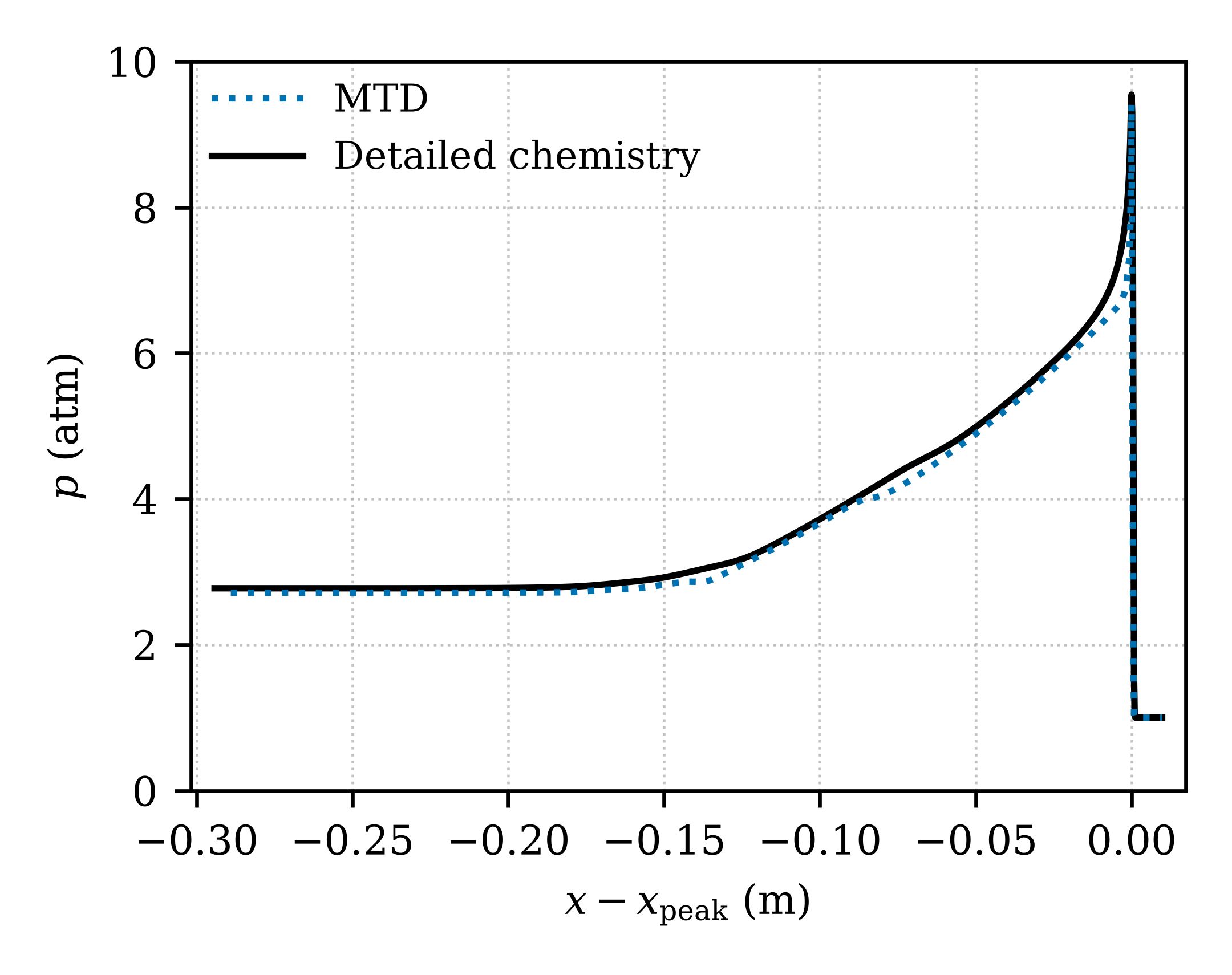}
    \vspace{-4mm}
    \caption{Pressure profile across 1D detonation for detailed chemistry and MTD solvers initiated from ignition kernel described in section \ref{sec:1d_det} for a stoichiometric $H_2$-air mixture.}
    \label{fig:p_compare_1d_det}
\end{figure}

\begin{figure}[H]
    \centering
    \includegraphics[width=0.5\linewidth]{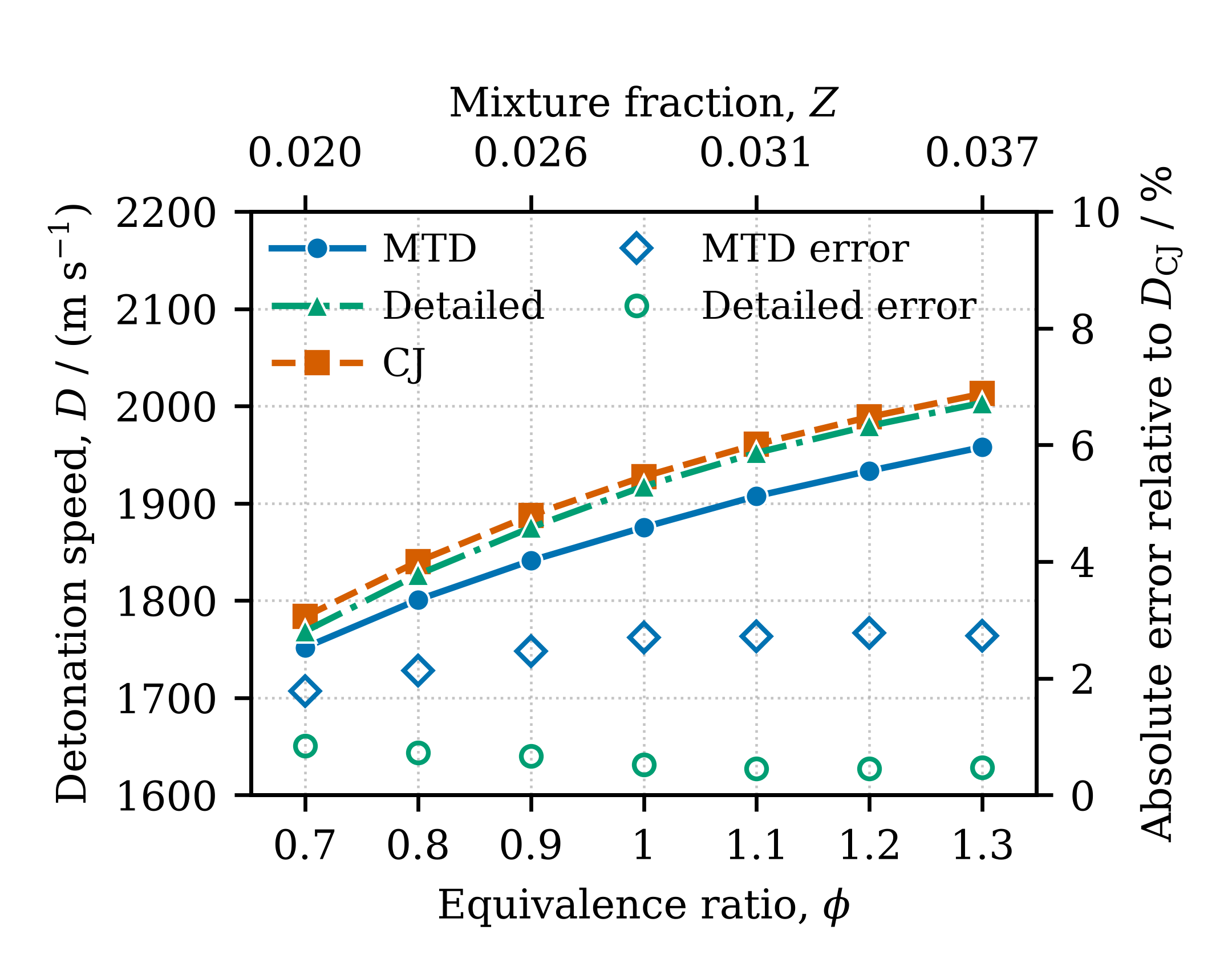}
    \vspace{-4mm}
    \caption{Measured detonation speed and relative error compared to CJ theory for detailed chemistry and MTD solvers initiated from ignition kernel described in section \ref{sec:1d_det} with respective mixture fractions.}
    \label{fig:wavespeed_compare_1d_det}
\end{figure}

The pressure profile of the detonation captured by the MTD solver is compared to the detailed chemistry solver in Fig.~\ref{fig:p_compare_1d_det}. There is a slight under-prediction of the von Neumann pressure and a sharper pressure drop behind the detonation front for the MTD solver compare to the detailed solver. The MTD post-detonation expansion profile shows good agreement with the detailed chemistry profile. Wave speed measurements of $w_{DC} = 1917.4 \hspace{1mm} m/s$ and $w_{MTD} = 1875.4 \hspace{1mm} m/s$ also show good agreement with the CJ speed for this case $1927.5 \hspace{1mm} m/s$. This results in a relative absolute percent error of under $3\%$ with respect to the CJ speed. This trend follows across mixture fractions as indicated in Fig.~\ref{fig:wavespeed_compare_1d_det} where all tested cases resulted in below $3\%$ relative absolute percent error. These results indicate that the MTD solver is capable of simulating a self-sustaining 1D detonation starting from initiation.

\vspace{-5mm}

\vspace{-4mm}

\subsection{2D RDE Reconstruction}

\begin{figure}[H]
    \centering
    \includegraphics[width=0.5\linewidth]{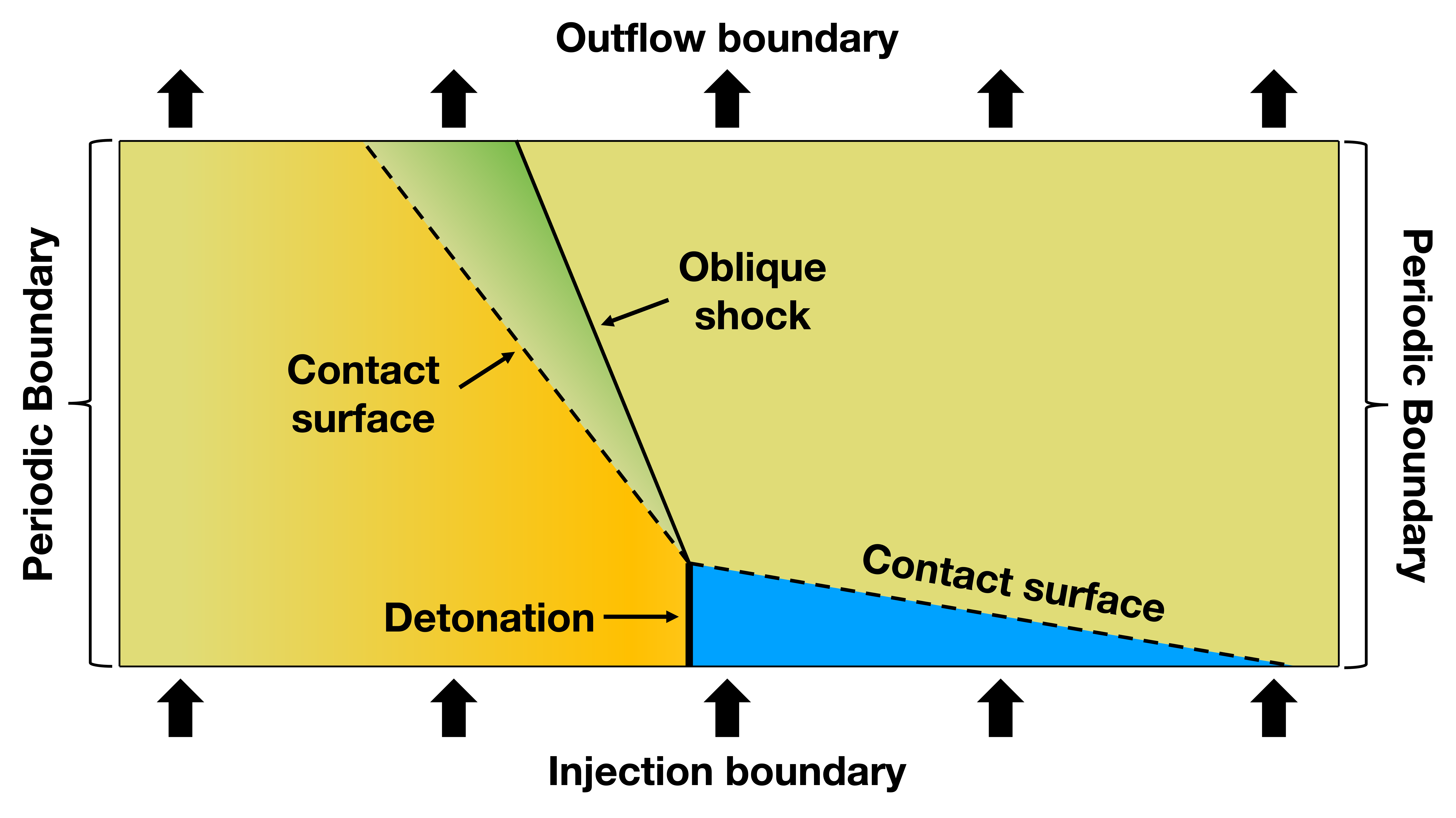}
    \vspace{-4mm}
    \caption{Schematic of 2D RDE geometry and boundary conditions adapted from \cite{wolanski_detonative_propulsion}.}
    \label{fig:domain_2d_rde}
\end{figure}

A richer problem of a 2D RDE is investigated which requires capturing more physics than just the chemistry in a 1D detonation. Figure~\ref{fig:domain_2d_rde} details the 2D RDE geometry and boundary conditions which includes a periodic boundary at the ends of the unwrapped circumference, an outflow boundary condition at the top, and a continuous injection boundary at the bottom. The domain has a height $H = 0.14 \hspace{1mm} m$ and length $L = 0.4788 \hspace{1mm} m$ and injects stoichiometric $H_2$-air with the plenum condition of $T_{0} = 300 \hspace{1mm} K$ and $p_{0} = 10 \hspace{1mm} atm$ (see \cite{ullman_2d_rde,sato_2d_rde} for details on the continuous injection boundary condition).

\begin{figure}[H]
    \centering
    \includegraphics[width=0.65\linewidth]{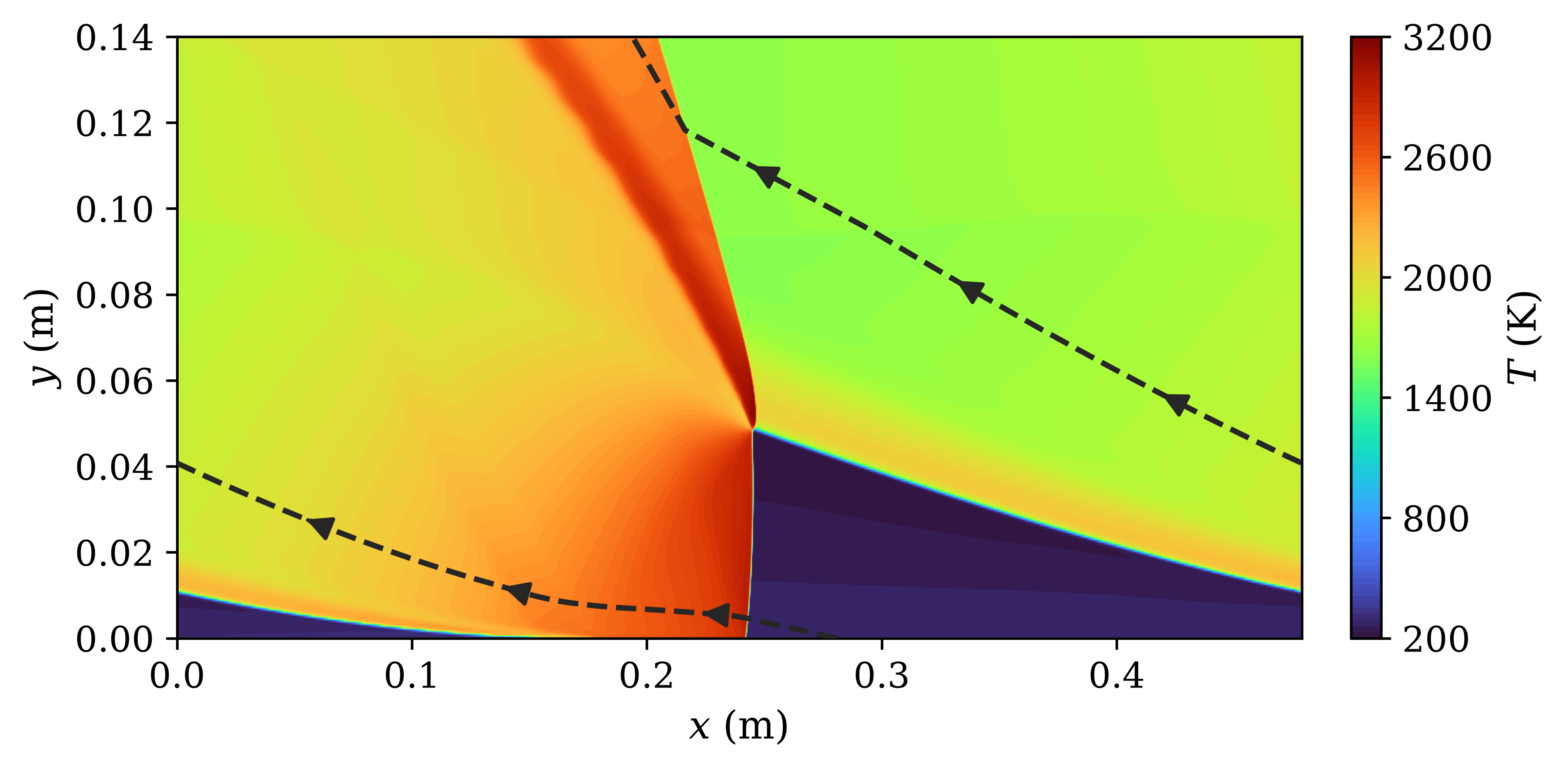}
    \vspace{-4mm}
    \caption{Instantaneous temperature contour from detailed chemistry case \cite{fernandez_2d_rde} with streamline corresponding to the extracted thermodynamic trajectory.}
    \label{fig:streamline_2d_rde}
\end{figure}

\begin{figure}[h]
    \centering
    \includegraphics[width=0.45\linewidth]{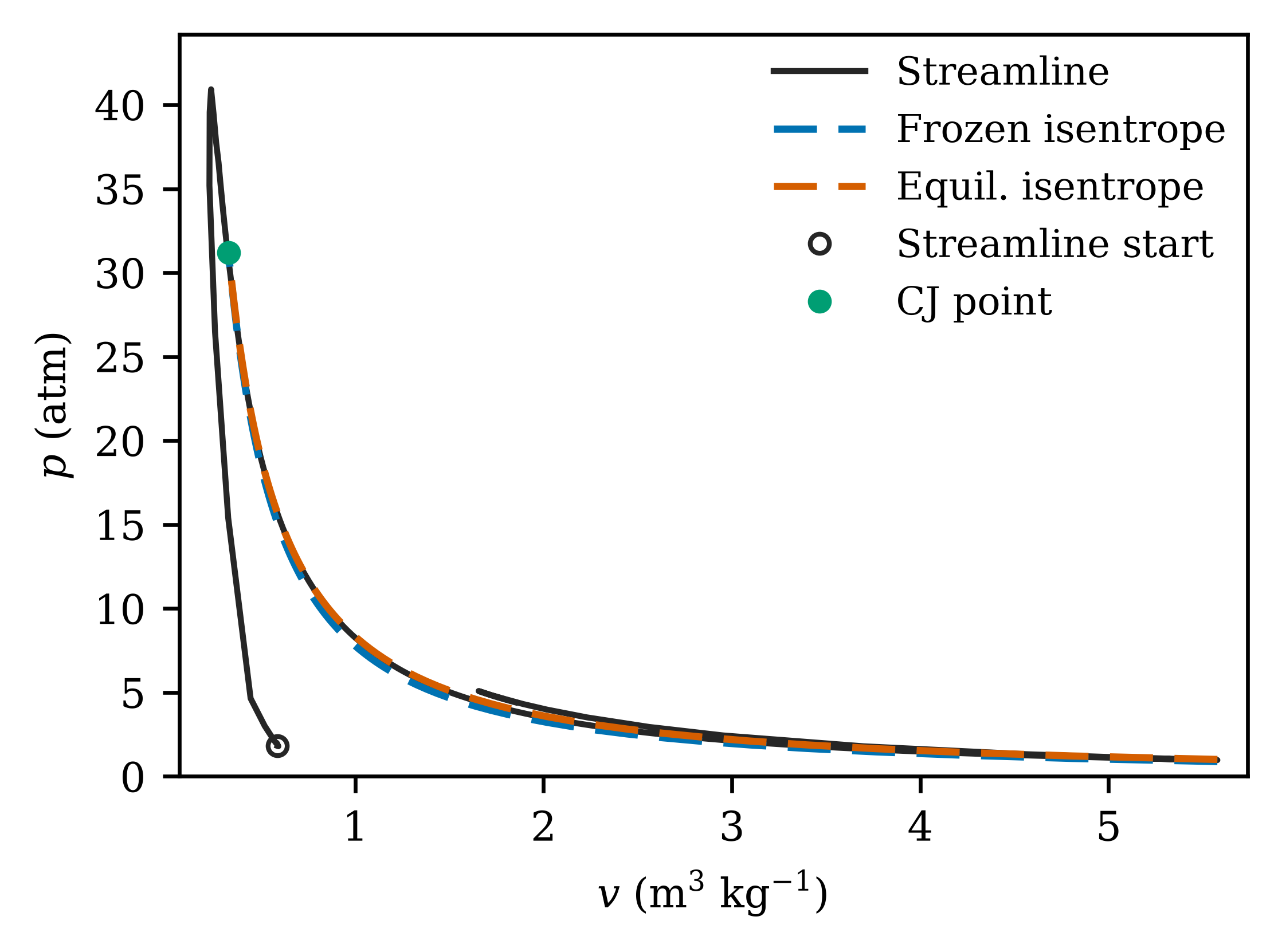}
    \vspace{-4mm}
    \caption{Thermodynamic trajectory along the streamline of a 2D RDE in $p-v$ space with the CJ point for the pre-detonation gas marked. Frozen and equilibrium isentropes are plotted from the CJ point alongside the actual trajectory of the streamline.}
    \label{fig:streamline_2d_rde_p_v}
\end{figure}

Results for the 2D RDE using the detailed chemistry solver \cite{fernandez_2d_rde} are used to extract the data from the streamline shown in Fig.~\ref{fig:streamline_2d_rde}. Of particular interest is the thermodynamic trajectory of the freshly injected gas along the streamline which is shown in Fig.~\ref{fig:streamline_2d_rde_p_v}. The injected gas expands before getting processed by the detonation which results in a higher pressure at the CJ point (equilibrium point behind a detonation) at an approximately equal specific volume with a much higher pressure. From the CJ point, the gas expands isentropically until it reaches the oblique shock attached to the triple point at the top of the detonation after which it expands until it exits the top boundary. This thermodynamic trajectory directly corresponds to the path of the streamline in Fig.~\ref{fig:streamline_2d_rde}. The previous discussion in section \ref{sec:expansion_model} is aided by this figure by the comparison of the equilibrium and frozen isentropes relative to the post-detonation expansion in the streamline data from the detailed chemistry solver.

\begin{figure}[h]
    \centering
    \includegraphics[width=0.45\linewidth]{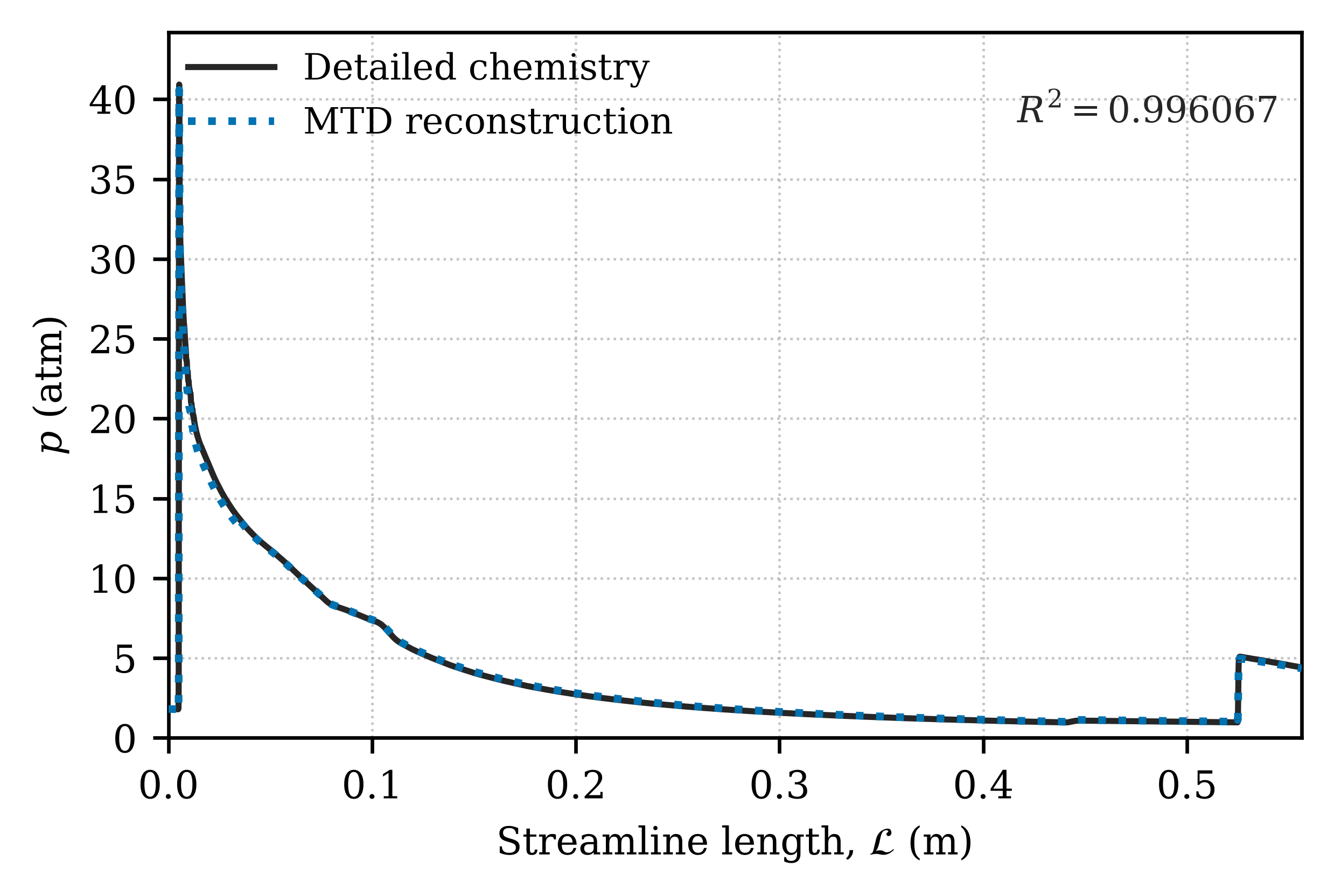}
    \caption{Pressure profile along streamline from 2D RDE data \cite{fernandez_2d_rde} and the MTD reconstruction.}
    \label{fig:streamline_2d_rde_reconstruction}
\end{figure}

The entire pressure profile along the streamline is reconstructed with the MTD model using only information from the conserved state in Fig.~\ref{fig:streamline_2d_rde_reconstruction}. All features are qualitatively captured in the MTD reconstruction: post-injection expansion, detonation, post-detonation expansion, and the post-detonation oblique shock. The fit of the reconstruction with respect to the streamline data is evaluated quantitatively via $R^2=0.996067$. This shows both qualitative and quantitative agreement with the streamline pressure profile.

\section{Conclusion}

A multi-mode tabulation framework is presented for deflagrative and detonative combustion, with applicability demonstrated across ignition, one-dimensional detonations, and two-dimensional rotating detonation engine (RDE) configurations. The framework replaces computationally expensive detailed chemistry integration with precomputed homogeneous reactor solutions for deflagrative combustion and post-detonation states up to the Chapman–Jouguet (CJ) point. Beyond this point, supersonic expansion is represented along an equilibrium isentrope originating from the CJ state, enabling a unified treatment of distinct combustion regimes.

The framework demonstrates promising accuracy across the test cases. For deflagrative combustion, the three-dimensional manifold reproduces detailed-chemistry ignition delays with a median deviation below 1\% at the reference pressure, although deviations at other pressures highlight the need for explicit pressure parameterization. In one-dimensional detonations, the model captures self-sustained wave initiation and propagation, with detonation speeds within 3\% of the CJ value across the tested mixture fractions and good agreement in post-detonation expansion. For the two-dimensional RDE, the framework reconstructs the complete streamline pressure evolution, including injection expansion, detonation, post-detonation expansion, and oblique shock interactions, directly from conserved flow variables, achieving an $R^2$ of 0.996 relative to detailed chemistry. Collectively, these results establish the potential of multi-mode tabulation to capture coupled combustion, shock, and expansion processes while reducing reliance on computationally expensive chemical kinetics integration in high-fidelity simulations of detonation-based propulsion systems.

Future work includes using MTD model to run the 2D RDE case from initiation and comparing the ability to capture the 2D RDE physcis as well as computational cost to the detailed chemistry solver.

\section*{CrediT authorship contribution statement}

\noindent NS: conceptualization, formal analysis, investigation, methodology, software, validation, visualization, writing—original draft, writing—review and editing\\
VS: conceptualization, formal analysis, investigation, methodology, software, validation, visualization, writing—review and editing\\
SS: conceptualization, formal analysis, investigation, software, validation, writing—review and editing \\
VR: conceptualization, formal analysis, investigation, funding acquisition, methodology, project administration, resources, supervision, writing—review and editing

\section*{Declaration of competing interest}

The authors declare that they have no known competing financial
interests or personal relationships that could have appeared to
influence the work reported in this paper.

\section*{Acknowledgments}

The research is supported by the Air Force Office of Scientific Research (AFOSR) under Grant No. FA9550-24-1-0017, with Dr. Chiping Li as Program Officer.

\section*{Appendix}

\subsection*{Z-C Transformation \label{app:derive}}

The model is presented using the governing equations for a multi-component reacting fluid

\begin{equation}
    \frac{\partial \rho}{\partial t} + \frac{\partial (\rho u)}{\partial x_j} = 0
\end{equation}

\begin{equation}
    \frac{\partial (\rho u_j)}{\partial t} + \frac{\partial (\rho u_i u_j)}{\partial x_j}+\frac{\partial p}{\partial x_j} - \frac{\partial \tau_{ij}}{\partial x_j} = 0
\end{equation}

\begin{equation}
    \frac{\partial (\rho E)}{\partial t}
    + \frac{\partial (\rho u_j E)}{\partial x_j}
    + \frac{\partial (u_j p)}{\partial x_j}
    - \frac{\partial (u_i \tau_{ij})}{\partial x_j}
    - \frac{\partial}{\partial x_j}
    \left[
        (\alpha + \rho D c_p) \frac{\partial T}{\partial x_j}
        - \rho D \frac{\partial h}{\partial x_j}
    \right] = 0
\end{equation}

\begin{equation}
    \frac{\partial (\rho Y_k)}{\partial t} + \frac{\partial (\rho u_j Y_k)}{\partial x_j} - \frac{\partial}{\partial x_j} \left( \rho D \frac{\partial Y_k}{\partial x_j} \right) = \Omega_k
\end{equation}

\noindent with an ideal gas equation of state where $\rho$ is the mixture density, $u_j$ is the velocity in direction $j$, $p$ is the thermodynamic pressure, $E$ is the total energy, and $Y_k$ is the mass fraction for species $k$ in a mixture with $N_s$ species such that $k \in [1,N_s]$. The viscous stress tensor $\tau_{ij}$ is defined by

\begin{align}
    \tau_{ij} = -\frac{2}{3} \mu \frac{\partial u_m}{\partial x_m} \delta_{ij} + \mu \left( \frac{\partial u_j}{\partial x_i} + \frac{\partial u_i}{\partial x_j} \right)
\end{align}

\noindent where $\mu$ is the dynamic viscosity. The heat flux $q$ is defined by $q = -\alpha \frac{\partial T}{\partial x_j}$, where $\alpha$ is the thermal conductivity and $T$ is the temperature. The enthalpy of each species $h_k$ is defined as the sum of the sensible and chemical components

\begin{align}
    h_k = \int_{T_0}^T c_{p,k} d\tau + \Delta h_{f,k}^0
    \label{eq:h_k}
\end{align}

\noindent where $c_{p,k}$ is the specific heat at constant pressure and $\Delta h_{j,k}^0$ is the heat of formation for each species such that the mixture enthalpy and specific heat at constant pressure are calculated using the relations $h = \sum_{k=1}^{N_s} h_k Y_k$ and $c_p=\sum_{k=1}^{N_s} c_{p,k} Y_k$, respectively.

Following the equations described above and in $\S$\ref{subsec:NSeq}, it is evident that information about the mixture is no longer available in Z-C space therefore a transformation is sought to remove dependence on mass fractions to solve the governing equations. The enthalpy-species diffusion term in the energy equation is the only term outside species transport that needs to be addressed. Using the species and mixture enthalpy equations (see equation \ref{eq:h_k}), the mixture enthalpy gradient can be written in terms of $c_{\text{p}}$

\begin{align}
    \frac{\partial h}{\partial x_j} &= \sum_{k=1}^{N_s} h_k \frac{\partial Y_k}{\partial x_j} + Y_k \frac{\partial h_k}{\partial x_j} \\
    &= \sum_{k=1}^{N_s} h_k \frac{\partial Y_k}{\partial x_j} + c_{p,k} Y_k \frac{\partial T}{\partial x_j} \\
    &= c_{\text{p}} \frac{\partial T}{\partial x_j} + \sum_{k=1}^{N_s} h_k \frac{\partial Y_k}{\partial x_j}
\end{align}

\noindent which is a tabulated property. This allows the transformed form of the energy equation to be written as

\begin{equation}
\begin{aligned}
    &\frac{\partial (\rho E)}{\partial t}
    + \frac{\partial (\rho u_j E)}{\partial x_j}
    + \frac{\partial (u_j p)}{\partial x_j} \\
    &- \frac{\partial (u_i \tau_{ij})}{\partial x_j}
    - \frac{\partial}{\partial x_j}
    \left[
        (\alpha + \rho D c_p) \frac{\partial T}{\partial x_j}
        - \rho D \frac{\partial h}{\partial x_j}
    \right] = 0
\end{aligned}
\end{equation}

\noindent where $\alpha$, $D$, and $c_p$ are all tabulated properties that can be evaluated for a given conserved state using the thermochemical closure mentioned in section \ref{sec:thermochemical_closure}. It then follows that the $Z$ and $C$ scalars in the transformed system have the corresponding transport equations

\begin{equation}
    \frac{\partial (\rho Z)}{\partial t} + \frac{\partial (\rho u_j Z)}{\partial x_j} - \frac{\partial}{\partial x_j} \left( \rho D \frac{\partial Z}{\partial x_j} \right) = 0
\end{equation}

\begin{equation}
    \frac{\partial (\rho C)}{\partial t} + \frac{\partial (\rho u_j C)}{\partial x_j} - \frac{\partial}{\partial x_j} \left( \rho D \frac{\partial C}{\partial x_j} \right) = \Omega_C
\end{equation}

\noindent where $\Omega_C$ is the tabulated progress variable production term.

\bibliographystyle{unsrtnat}
\bibliography{references}
\end{document}